\documentclass{article}
\usepackage{spconf,amsmath,graphicx,setspace}
\usepackage{amssymb}
\usepackage{algorithm}
\usepackage{algorithmic}
\usepackage{textcomp}
\usepackage{subfigure}
\usepackage{tipa}
\usepackage{cite} 
\usepackage{url}
\usepackage{float}

\makeatletter 
\renewcommand{\@thesubfigure}{\hskip\subfiglabelskip}
\makeatother
\title{SELF-KNOWLEDGE DISTILLATION BASED SELF-SUPERVISED LEARNING FOR COVID-19 DETECTION FROM CHEST X-RAY IMAGES}
\name{Guang Li $^{\dagger}$ \qquad Ren Togo $^{\dagger\dagger}$ \qquad Takahiro Ogawa $^{\dagger\dagger\dagger}$ \qquad Miki Haseyama$^{\dagger\dagger\dagger}$ \thanks{
This study was supported in part by AMED Grant Number JP21zf0127004. This  study  was  conducted  on  the  Data  Science  Computing System of Education and Research Center for Mathematical and Data Science, Hokkaido University.}}
\address{$^{\dagger}$ Graduate School of Information Science and Technology,
    Hokkaido University, Japan \\
    $^{\dagger\dagger}$ Education and Research Center for Mathematical and Data Science,
 	Hokkaido University, Japan \\
    $^{\dagger\dagger\dagger}$ Faculty of Information Science and Technology, 
    Hokkaido University, Japan \\
 	E-mail: \{guang, togo, ogawa\}@lmd.ist.hokudai.ac.jp, miki@ist.hokudai.ac.jp}
\begin{document}
\ninept
\maketitle
%
\begin{abstract}
\end{abstract}
The global outbreak of the Coronavirus 2019 (COVID-19) has overloaded worldwide healthcare systems.
Computer-aided diagnosis for COVID-19 fast detection and patient triage is becoming critical. 
This paper proposes a novel self-knowledge distillation based self-supervised learning method for COVID-19 detection from chest X-ray images.
Our method can use self-knowledge of images based on similarities of their visual features for self-supervised learning.
Experimental results show that our method achieved an HM score of 0.988, an AUC of 0.999, and an accuracy of 0.957 on the largest open COVID-19 chest X-ray dataset.
\par
\section{Introduction}
The Coronavirus 2019 (COVID-19) caused by Severe Acute Respiratory Syndrome Coronavirus 2 (SARS-CoV2) has rapidly spread worldwide, which is declared a global pandemic~\cite{andersen2020proximal}.
As of 16 September 2021, there have been 226,236,577 confirmed cases of COVID-19, including 4,654,548 deaths worldwide~\footnote{https://covid19.who.int}.
In the light of the unprecedented pandemic by COVID-19, public healthcare systems have faced many challenges such as scarce medical resources, which are pushing healthcare providers to face the threat of infection~\cite{liu2020experiences}.
Considering the ominously contagious nature of COVID-19, the early screening for COVID-19 has become increasingly important to prevent the further spread of the disease and reduce the burden on saturated healthcare providers.
\par
Real-time Reverse Transcription Polymerase Chain Reaction (RT-PCR) is currently considered the gold standard for COVID-19 detection~\cite{pujadas2020comparison}.
However, RT-PCR is reported with a high false-negative rate and is time-consuming, depriving health authorities of the opportunity of early isolation~\cite{drame2020should}. 
Because radiological findings of pneumonia are commonly presented in COVID-19 patients, radiologic examinations are often used for diagnosing COVID-19~\cite{shi2020radiological}.
In particular, chest X-ray has advantages in terms of the low cost, short scan time, and a low dose of radiation compared to Computed Tomography (CT)~\cite{huang2021graph}.
Therefore, computer-aided chest radiography diagnosis has a great potential for the fast screening of COVID-19 and analysis of patients' conditions, which may help radiologists under global pandemic~\cite{oh2020deep}.
\par
Self-supervised learning has recently attracted widespread attention in machine learning~\cite{liu2020self}.
Unlike supervised learning using manually annotated labels, self-supervised learning benefits from image characteristics ($e.g.$, texture, position, and color) without manually annotated labels~\cite{jing2020self}.
For example, the study~\cite{noroozi2016unsupervised} plays a jigsaw game on images, and the study~\cite{gidaris2018unsupervised} predicts the rotation degrees of images.
Particularly, self-supervised learning methods based on the Siamese network have been shown to be effective on different medical datasets~\cite{li2021triplet}.
These methods define the inputs as two augmented views from one image, then input to the Siamese network and maximize the similarity between the representations of views~\cite{li2021cross}.
Self-supervised learning is considered helpful for medical image processing, and we introduce the self-knowledge distillation for learning better representations of COVID-19 chest X-ray images. 
\par
In this paper, we propose a novel self-knowledge distillation based self-supervised learning method for COVID-19 detection from chest X-ray images.
Our method can make use of self-knowledge of images based on similarities of their visual features for self-supervised learning.
Experimental results show that we can achieve an HM score of 0.988, an AUC of 0.999, and an accuracy of 0.957 on the largest open COVID-19 chest X-ray dataset~\cite{rahman2021exploring}.
Our method may help to prevent the further spread of the COVID-19 and reduce the burden on healthcare providers and radiologists.
\begin{figure*}[t]
        \centering
        \includegraphics[width=17cm]{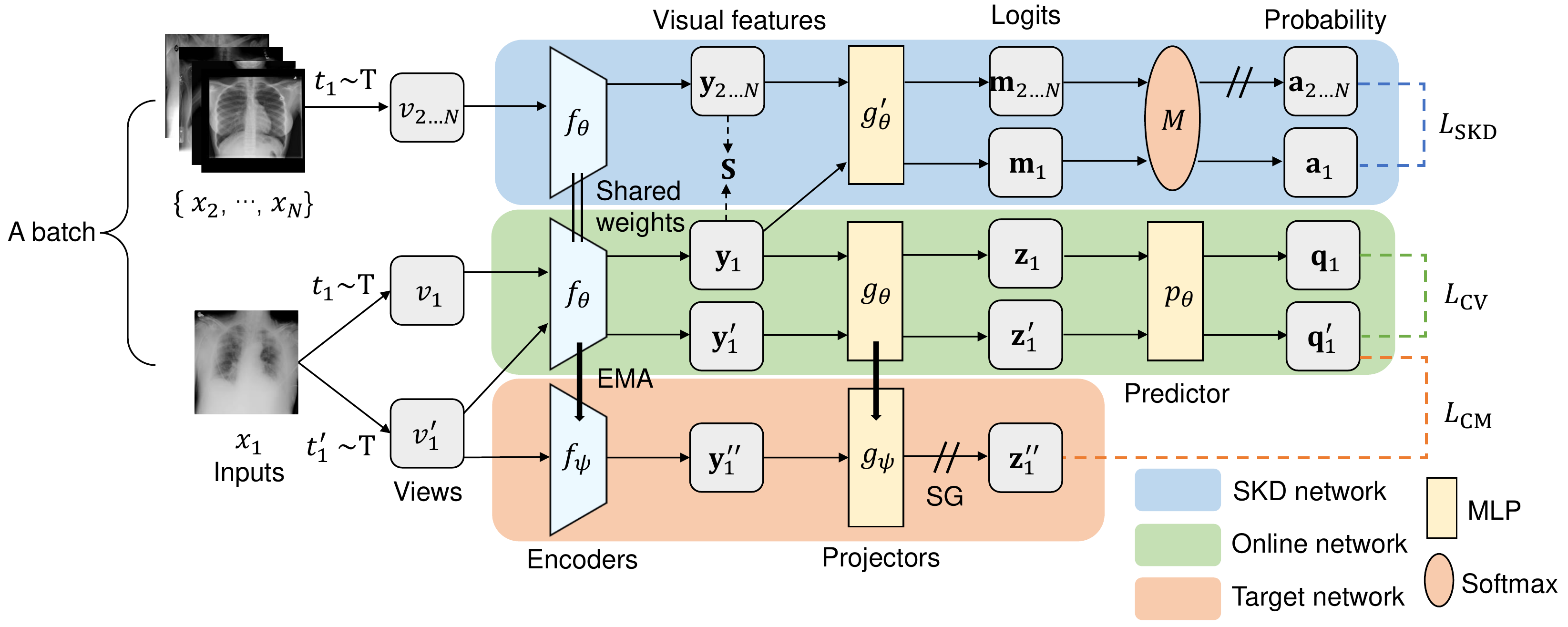}
        \caption{Overview of the proposed method. The weights of the target network ($\psi$) are an exponential moving average (EMA) of the weights of the online network ($\theta$), and the encoders ($f_{\theta}$) in self-knowledge distillation network (SKD network) and online network share the weights. MLP denotes multilayer perceptron. SG denotes stop-gradient.}
        \label{fig1}
\end{figure*}
\par
Our contributions are summarized as follows.
\begin{itemize}
    \item We propose a novel self-knowledge distillation based self-supervised learning method for COVID-19 detection from chest X-ray images.
    \item We realize promising detection results on the largest open COVID-19 chest X-ray dataset.
\end{itemize}
\section{Self-Knowledge Distillation based Self-Supervised Learning}
Our method uses triplet networks (a variant of the Siamese network)~\cite{li2022tri} to learn discriminative representations from chest X-ray images.
Figure~\ref{fig1} shows an overview of the proposed method.
The proposed method comprises three networks, where the weights of the target network are an exponential moving average (EMA) of the weights of the online network, and the encoders in self-knowledge distillation network (SKD network) and online network share the weights~\cite{antti2017mean}.
Our method consists of two modules, a self-supervised learning module and a self-knowledge distillation module.
We show the details of the proposed method in the following subsections.
\subsection{Self-Supervised Learning Module}
First, we introduce the self-supervised learning module of our method.
Given an input chest X-ray image $x_{1}$ in a batch  of $N$ images, two transformations $t_{1}$ and $t'_{1}$ are randomly sampled from the distribution $T$ to generate two views $v_{1} = t_{1}(x_{1})$ and $v'_{1} = t'_{1}(x_{1})$.
Specifically, these transformations combine standard data augmentation methods such as cropping, resizing, flipping, and Gaussian blur as described in~\cite{tian2020makes}.
The view $v_{1}$ is processed by the encoder $f_{\theta}$ and projector $g_{\theta}$ of the online network.
Accordingly, the view $v'_{1}$ is processed by the encoder $f_{\psi}$ and projector $g_{\psi}$ of the target network ($\mathbf{z}''_{1}$ is the final output of $v'_{1}$).
Specifically, we obtain a copy of $v'_{1}$ and input it into the online network to calculate the final loss.
Furthermore, we use predictor $p_{\theta}$ to transform the final outputs of two views as $\mathbf{q}_{1}$ and $\mathbf{q}'_{1}$ in the online network.
\par
We define the similarity losses between normalized predictions and projections.
The cross-view loss $L_{\mathrm{CV}}$ compares the representations of two views from the online network and is defined as follows:
\begin{equation}
\begin{split}
L_{\mathrm{CV}} 
& = || \hat{\mathbf{q}}_{1} - \hat{\mathbf{q}}'_{1} ||_{2}^{2}
\\ & = 2 - 2 \cdot \frac{\left \langle \mathbf{q}_{1},\mathbf{q}'_{1} \right \rangle}{ || \mathbf{q}_{1} ||_{2} \cdot || \mathbf{q}'_{1} ||_{2}},
\end{split}
\end{equation}
where $\hat{\mathbf{q}}_{1} = \mathbf{q}_{1}/ || \mathbf{q}_{1} ||_{2}$ and $\hat{\mathbf{q}}'_{1} = \mathbf{q}'_{1}/ || \mathbf{q}'_{1} ||_{2}$ denote the normalized predictions of $v_{1}$ and $v'_{1}$ from the online network, respectively.
The cross-model loss $L_{\mathrm{CM}}$ compares the representations of the same view from the online network and the target network and is defined as follows:
\begin{equation}
\begin{split}
L_{\mathrm{CM}} 
& = || \hat{\mathbf{q}}'_{1} - \hat{\mathbf{z}}''_{1} ||_{2}^{2} 
\\ & = 2 - 2 \cdot \frac{\left \langle \mathbf{q}'_{1}, \mathbf{z}''_{1} \right \rangle}{ || \mathbf{q}'_{1} ||_{2} \cdot || \mathbf{z}''_{1} ||_{2}},
\end{split}
\end{equation}
where $\hat{\mathbf{z}}''_{1} = \mathbf{z}''_{1}/ || \mathbf{z}''_{1} ||_{2}$ denotes the normalized projection of $v'_{1}$ from the target network.
Note that we only use the predictor in the online network to make the architecture asymmetric, preventing learning from collapsing~\cite{grill2020bootstrap}.
We can learn good representations from each chest X-ray image alone with the cross-view loss and the cross-model loss.
\begin{figure*}[t]
        \centering
        \subfigure[(a)]{
        \centering
        \includegraphics[width=4.0cm]{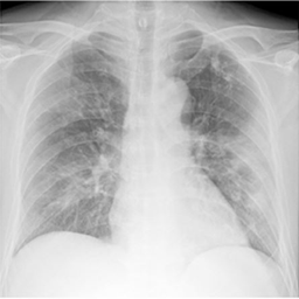}
        }
        \subfigure[(b)]{
        \centering
        \includegraphics[width=4.0cm]{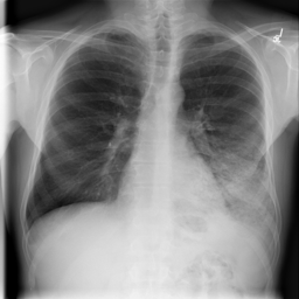}
        }
        \subfigure[(c)]{
        \centering
        \includegraphics[width=4.0cm]{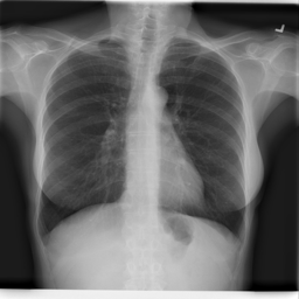}
        }
        \subfigure[(d)]{
        \centering
        \includegraphics[width=4.0cm]{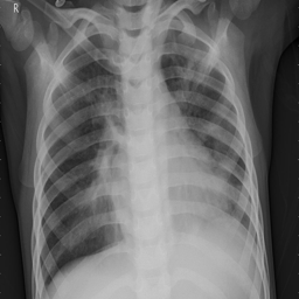}
        }
        \caption{Examples of chest X-ray images: (a) COVID-19 , (b) Lung Opacity (c) Normal, and (d) Viral Pneumonia.}
        \label{fig2}
\end{figure*}
\subsection{Self-Knowledge Distillation Module}
Next, we introduce the self-knowledge distillation module of our method.
Self-knowledge distillation can be considered as regularizing the training of a network using soft targets that carry the ``dark knowledge" of the same network.
Since images with highly similar visual features tend to have similar predicted probabilities, similar images' knowledge could be ensembled to provide better soft targets for self-knowledge distillation.   
With the above concept, we can learn better representations from different chest X-ray images based on the similarities of their visual features.
As shown in Fig.~\ref{fig1}, we first obtain the similarity matrix $\mathbf{S} \in \mathbb{R}^{N \times N}$ by comparing the encoded visual features $\{\mathbf{y}_{1}, ..., \mathbf{y}_{N}\}$ in a batch of $N$ images as follows:
\begin{equation}
\mathbf{S}_{i,j} = (\hat{\mathbf{y}}_{i}^{\top} \hat{\mathbf{y}}_{j}),
\end{equation}
where $\hat{\mathbf{y}}_{i} = \mathbf{y}_{i}/ || \mathbf{y}_{i} ||_{2}$ denotes the normalized representations, $i,j$ are the indices in the batch, and we discard the diagonal entries with an identity matrix $\mathbf{I}$ to avoid self-knowledge reinforcement by $ \mathbf{S} = \mathbf{S} \odot (1-\mathbf{I})$.
Subsequently, we normalize the similarity matrix as follows:
\begin{equation}
\hat{\mathbf{S}}_{i,j} = \frac{\mathrm{exp}(\mathbf{S}_{i,j})}{\sum_{j \neq i}\mathrm{exp}(\mathbf{S}_{i, j})}, \forall i \in \{1,...,N\}.
\end{equation}
The predictive probabilities $\{\mathbf{a}_{1}, ..., \mathbf{a}_{N}\}$ can be obtained by $g'_{\theta}$ and a softmax function $M$ of the SKD network on the output logits $\{\mathbf{m}_{1}, ..., \mathbf{m}_{N}\}$ of the SKD network as follows:
\begin{equation}
\mathbf{a}_{(k)} = \frac{\mathrm{exp}(\mathbf{m}_{k}/\tau)}{\sum_{i = 1}^{K}\mathrm{exp}(\mathbf{m}_{i}/\tau)},
\end{equation}
where $K$ denotes the dimensions of output logits, and $\tau$ is a temperature hyperparameter. 
We denote the predicted probabilities of images within a batch as $\mathbf{A} = [\mathbf{a}_{1}, ..., \mathbf{a}_{N}]^{\top} \in \mathbb{R}^{N \times K}$.
To avoid propagating and ensembling noisy predictions too much, we generate the soft targets $\mathbf{B}$ as a weighted sum of the initial probability matrix $\mathbf{A}$ and the propagated probability matrix $\hat{\mathbf{S}}\mathbf{A}$ as follows:
\begin{equation}
\mathbf{B} = \omega\hat{\mathbf{S}}\mathbf{A} + (1-\omega)\mathbf{A}.
\end{equation}
And we can propagate multiple times to generate better soft targets $\mathbf{B}$ for self-knowledge distillation as follows:
\begin{equation}
\begin{split}
\mathbf{B}_{(t)}
& = \omega\hat{\mathbf{S}}\mathbf{B}_{(t-1)} + (1-\omega)\mathbf{A},
\\ & = (\omega\hat{\mathbf{S}})^{t}\mathbf{A} + (1-\omega)\sum_{i = 0}^{t-1}(\omega\hat{\mathbf{S}})^{i}\mathbf{A},
\end{split}
\end{equation}
where $\omega$ is a weight factor, and $t$ denotes $t$-th propagation and ensembling iteration.
When the number of iterations approaches infinite, we have $\mathrm{lim}_{t\rightarrow\infty}(\omega\hat{\mathbf{S}})^{t} = 0$ and $\mathrm{lim}_{t\rightarrow\infty}\sum_{i = 0}^{t-1}(\omega\hat{\mathbf{S}})^{i} = (\mathbf{I} - \omega\hat{\mathbf{S}})^{-1}$, and hence we can obtain an approximate inference formulation as follows:
\begin{equation}
\mathbf{B} = (1-\omega)(\mathbf{I} - \omega\hat{\mathbf{S}})^{-1}\mathbf{A}.
\end{equation}
Finally, we can define the self-knowledge distillation loss $L_{\mathrm{SKD}}$ as follows:
\begin{equation}
L_{\mathrm{SKD}} = D_\mathrm{KL}(\mathbf{B}||\mathbf{A}),
\end{equation}
where $D_\mathrm{KL}$ denotes KL divergence.
With the self-knowledge distillation loss, we can learn better representations from different chest X-ray images based on the similarities of their visual features.
\subsection{Network Optimization}
Finally, we show how to optimize the networks of our method.
The weights of the online network ($\theta$) are updated by minimizing a total loss $L$ that combines the similarity losses and self-knowledge distillation loss as follows:
\begin{equation}
L = L_{\mathrm{CV}} + L_{\mathrm{CM}} + \lambda \cdot \tau^{2} \cdot L_{\mathrm{SKD}},
\end{equation}
\begin{equation}
\theta \leftarrow \mathrm{Opt}(\theta, \nabla_{\theta}L, \alpha),
\end{equation}
where $\lambda$ is a balance hyperparameter, $\mathrm{Opt}$ and $\alpha$ denote the optimizer and the learning rate, respectively.
The weights of the target network ($\psi$) are an exponential moving average of the weights of the online network ($\theta$) and are updated as follows:
\begin{equation}
\psi \leftarrow \sigma\psi + (1-\sigma)\theta,
\end{equation}
where $\sigma$ denotes the degree of moving average, and we update weights after every iteration.
The gradient is not back-propagated through the soft targets and the target network for stable training~\cite{chen2021exploring}. 
After the self-knowledge distillation based self-supervised learning, the encoder of the online network ($f_{\theta}$) can learn discriminative representations from chest X-ray images and can be used for fine-tuning and COVID-19 detection.
\section{Experiments}
\subsection{Dataset and Settings}
\begin{table}[t]
    \centering
    \caption{Details of the COVID-19 chest X-ray dataset~\cite{rahman2021exploring} used in our study.}
    \label{tab1}
    \begin{tabular}{lccc}
    \hline
    Class & Total & Training image & Test image\\\hline
    COVID-19
    & 3,616 & 2,893 & 723 \\
    Lung Opacity
    & 6,012 & 4,810 & 1,202 \\
    Normal
    & 10,192 & 8,154 & 2,038 \\
    Viral Pneumonia
    & 1,345 & 1,076 & 269 \\
    \hline
    \end{tabular}
\end{table}
The dataset used in our research is the largest open COVID-19 chest X-ray dataset~\cite{rahman2021exploring}.
As shown in Table~\ref{tab1}, the dataset has 4 classes ($i.e.$, COVID-19, Lung Opacity, Normal, and Viral Pneumonia) with a total number of 21,165 images.
We randomly select 80\% chest X-ray images as the training set and the remaining 20\% as the test set.
Figure~\ref{fig2} shows examples of chest X-ray images.
All of these images are grayscale and have a resolution of 224 $\times$ 224 pixels.
We used sensitivity (Sen), specificity (Spe), their harmonic mean (HM), area under the ROC curve (AUC), and four-class classification accuracy (Acc) as evaluation metrics.
For Sen, Spe, HM, and AUC, we took COVID-19 as positive and the others as negative.
\par
We used the ResNet50~\cite{he2016deep} network as our encoders $f_{\theta}$ and $f_{\psi}$, whose output feature dimension was 2,048 (the output of the final average pooling layer).
The optimizer used in our method was an SGD optimizer, whose learning rate $\alpha$, momentum, and weight decay were set to 0.03, 0.9, 0.0004, respectively.
The projectors $g_{\theta}$ and $g_{\psi}$ and predictor $p_{\theta}$ are multilayer perceptron (MLP) with the same architecture, comprising a linear layer with an output size of 4,096, a batch normalization layer, a ReLU activation function, and a linear layer with an output size of 256~\cite{grill2020bootstrap}.
We took self-supervised learning on the dataset for 40 epochs and fine-tuning on the dataset for 30 epochs.
The batch size was 256, and generated view size was 112. 
Hyperparameters $\lambda$, $\tau$, $\omega$, $K$, and $\sigma$ were set to 1.0, 4.0, 0.5, 4.0, and 0.996, respectively~\cite{grill2020bootstrap, ge2021self}.
\par
\begin{table*}[t]
    \centering
    \caption{Test results of COVID-19 detection (average of the last 10 fine-tuning epochs).}
    \label{tab2}
    \begin{tabular}{lccccc}
    \hline
    Method & Sen & Spe & HM & AUC & Acc \\\hline
    Ours
    & \bfseries{0.980$\pm$0.004} & \bfseries{0.997$\pm$0.001} & \bfseries{0.988$\pm$0.002} & \bfseries{0.999$\pm$0.000} & \bfseries{0.957$\pm$0.001} \\
    Cross~\cite{li2021self}
    & 0.972$\pm$0.003 & 0.997$\pm$0.001 & 0.985$\pm$0.001 & 0.999$\pm$0.000 & 0.953$\pm$0.001 \\
    BYOL~\cite{grill2020bootstrap}
    & 0.973$\pm$0.004 & 0.996$\pm$0.001 & 0.985$\pm$0.002 & 0.999$\pm$0.000 & 0.954$\pm$0.001 \\
    SimSiam~\cite{chen2021exploring} 
    & 0.974$\pm$0.004 & 0.995$\pm$0.001 & 0.984$\pm$0.002 & 0.998$\pm$0.000 & 0.950$\pm$0.001 \\
    PIRL-Jigsaw~\cite{misra2020self}
    & 0.977$\pm$0.003 & 0.997$\pm$0.001 & 0.987$\pm$0.001 & 0.999$\pm$0.000 & 0.951$\pm$0.001 \\
    PIRL-Rotation~\cite{misra2020self}
    & 0.973$\pm$0.002 & 0.997$\pm$0.001 & 0.985$\pm$0.001 & 0.999$\pm$0.000 & 0.951$\pm$0.001 \\
    SimCLR~\cite{chen2020simple}
    & 0.913$\pm$0.006  & 0.994$\pm$0.001 & 0.952$\pm$0.003 & 0.996$\pm$0.000 & 0.936$\pm$0.001 \\
    Transfer
    & 0.944$\pm$0.004 & 0.994$\pm$0.001 & 0.968$\pm$0.002 & 0.997$\pm$0.000 & 0.936$\pm$0.001 \\
    From Scratch
    & 0.665$\pm$0.013 & 0.954$\pm$0.003 & 0.783$\pm$0.008 & 0.935$\pm$0.001 & 0.774$\pm$0.002 \\
    \hline
    \end{tabular}
\end{table*}
\begin{table}[t]
    \centering
    \caption{COVID-19 detection results reported in~\cite{rahman2021exploring}.}
    \label{tab3}
    \begin{tabular}{lcccc}
    \hline
    Method & Sen & Spe & HM & Acc \\\hline
    Ours (ResNet50)
    & \bfseries{0.980} & \bfseries{0.997} & \bfseries{0.988} & \bfseries{0.957} \\
    ResNet18
    & 0.934 & 0.955 & 0.944 & 0.934 \\
    ResNet50
    & 0.930 & 0.955 & 0.942  & 0.930 \\
    ResNet101
    & 0.930 & 0.951 & 0.940 & 0.930 \\
    ChexNet
    & 0.932 & 0.955 & 0.943 & 0.932 \\
    DenseNet201
    & 0.927 & 0.954 & 0.940 & 0.927 \\
    InceptionV3
    & 0.935 & 0.955 & 0.945 & 0.935 \\
    \hline
    \end{tabular}
\end{table}
\begin{figure}[t]
        \centering
        \subfigure[(a)]{
        \centering
        \includegraphics[width=4cm]{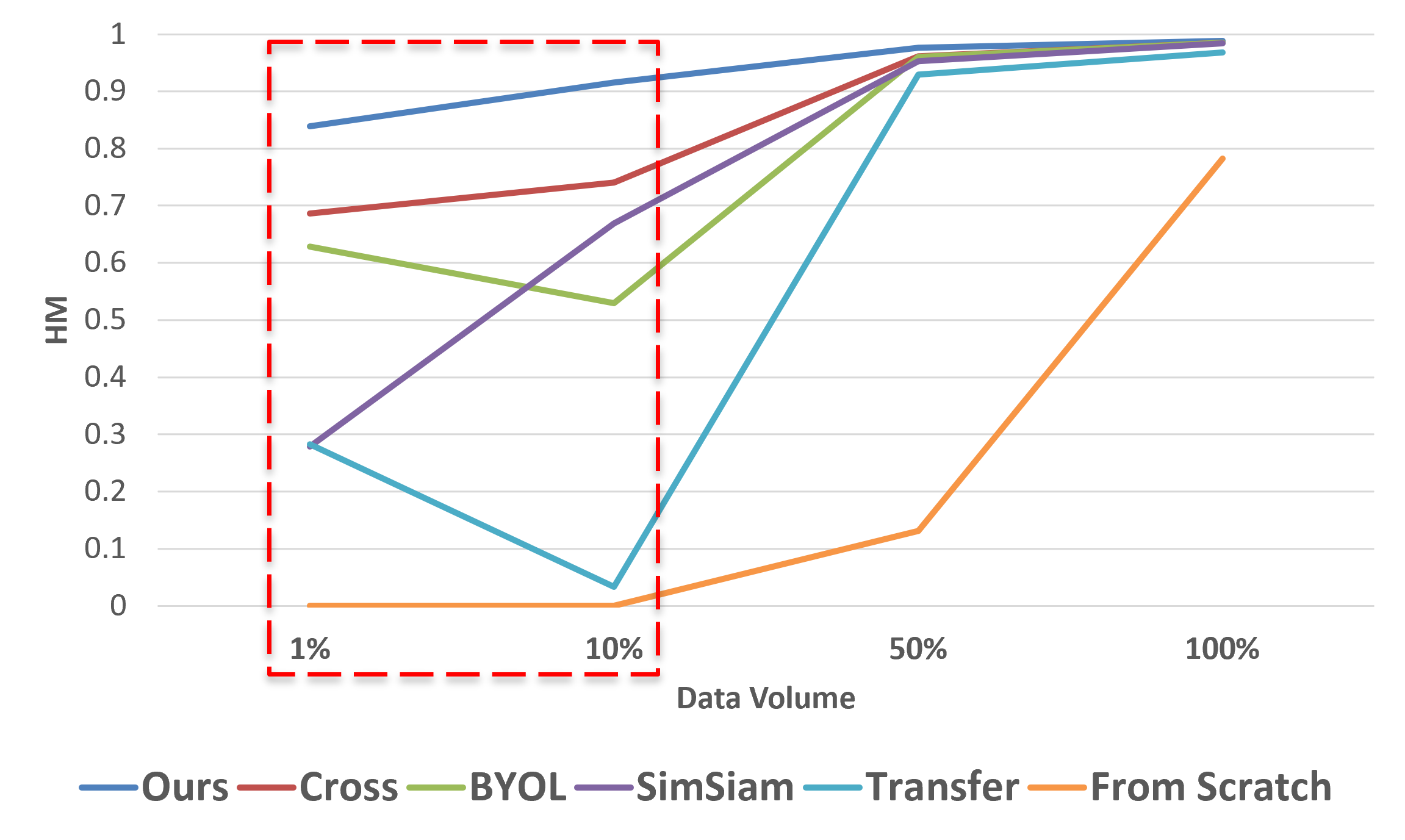}
        }
        \subfigure[(b)]{
        \centering
        \includegraphics[width=4cm]{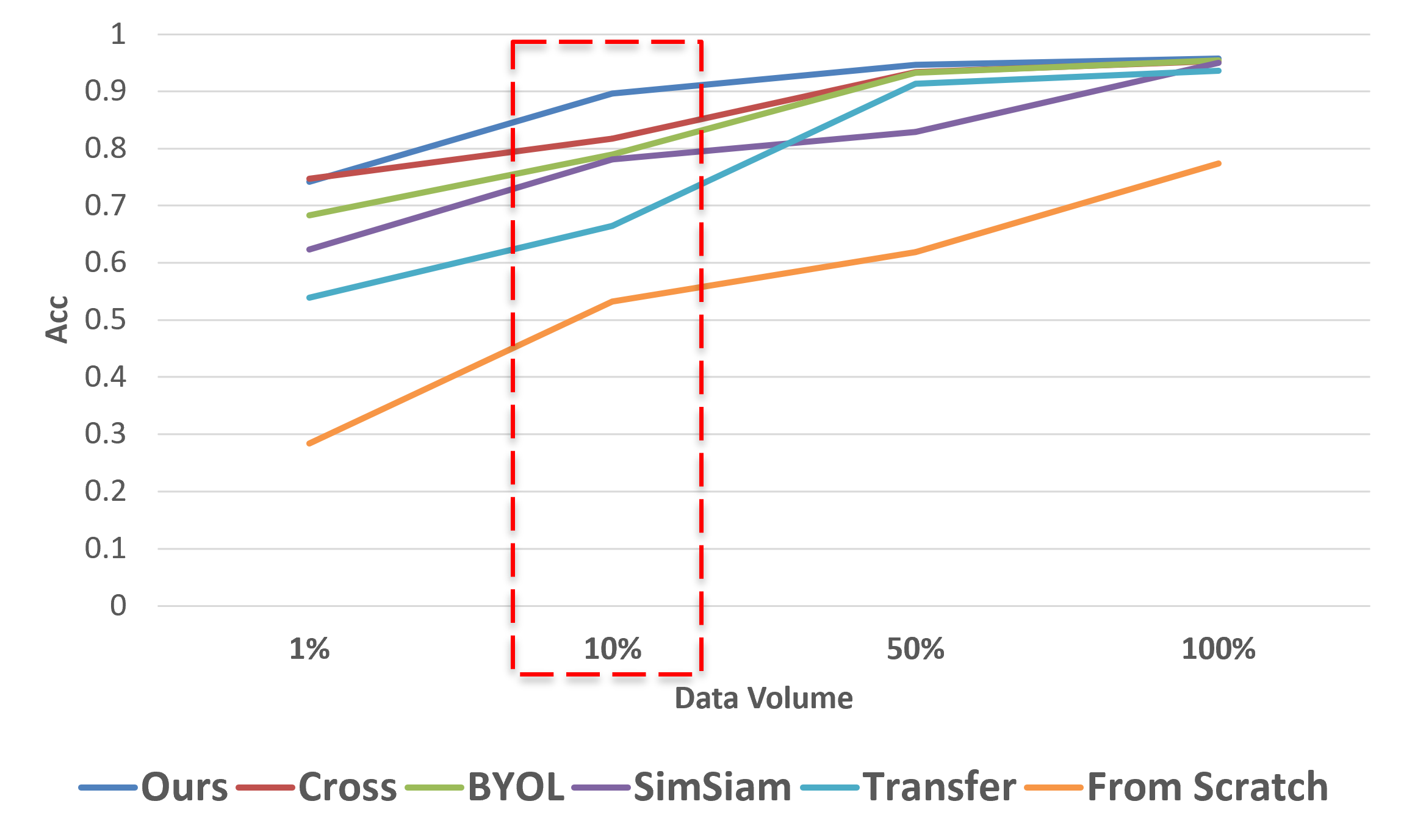}
        }
        \caption{Test results of COVID-19 detection in different data volumes: (a) HM and (b) Accuracy.}
        \label{fig3}
\end{figure}
\begin{figure}[t]
        \centering
        \includegraphics[width=8cm]{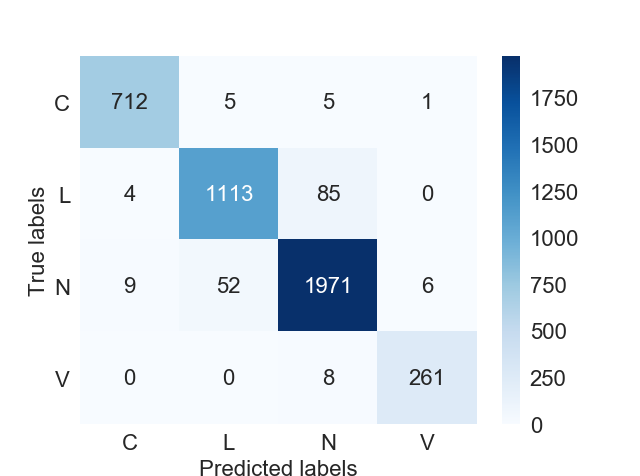}
        \caption{Confusion matrix for the best model of our method with an HM score of 0.990, an AUC of 1.000, and an accuracy of 0.959. ``C": COVID-19, ``L": Lung Opacity, ``N": Normal, and ``V": Viral Pneumonia.}
        \label{fig4}
\end{figure}
For comparative methods, we used several state-of-the-art self-supervised learning methods such as our self-supervised learning module alone (Cross)~\cite{li2021self}, BYOL~\cite{grill2020bootstrap}, SimSiam~\cite{chen2021exploring}, PIRL~\cite{misra2020self}, and SimCLR~\cite{chen2020simple}.
We also used Transfer learning (using ImageNet~\cite{deng2009imagenet} pre-trained weights) and training from scratch as baseline methods.
To verify the effectiveness of our method with a small amount of data, we randomly selected objects at 1\%, 10\%, 50\%, and 100\% of the training set size for the fine-tuning process and COVID-19 detection.
\subsection{Results and Discussion}
Test results of COVID-19 detection are shown in Table~\ref{tab2}.
We can see that our method outperformed all state-of-the-art self-supervised learning methods.
Moreover, the proposed method outperformed the self-supervised learning module alone (Cross), which shows the effectiveness of our self-knowledge distillation module.
Table~\ref{tab3} shows COVID-19 detection results reported in~\cite{rahman2021exploring}.
We can see that the proposed method that uses ResNet50 as the backbone drastically outperformed other methods.
Figure~\ref{fig3} shows the COVID-19 detection results in different data volumes.
From Fig.~\ref{fig3}, we can see that our method can significantly improve COVID-19 detection performance in different data volumes compared to other methods, and our method can achieve promising detection performance even using only 10\% data of the training set.
Furthermore, Fig.~\ref{fig4} shows the confusion matrix for the best model of our method with an HM score of 0.990, an AUC of 1.000, and an accuracy of 0.959.
\par
Considering the large number of patients who are being screened due to the global pandemic of COVID-19, the use of deep learning for computer-aided diagnosis has strong potential to assist in clinical workflow efficiency and reduce the burden on healthcare providers and radiologists.
Our findings show the effectiveness of self-knowledge distillation based self-supervised learning for COVID-19 detection from chest X-ray images.
Although the experimental results are promising, the proposed method should be evaluated on other COVID-19 chest X-ray image datasets for any potential bias.
\section{Conclusion}
We have proposed a novel self-knowledge distillation based self-supervised learning method for COVID-19 detection from chest X-ray images.
Our method can make use of self-knowledge of images based on similarities of their visual features for self-supervised learning.
Experimental results show that we realize high detection performance on the largest open COVID-19 chest X-ray dataset.
Our method may help to prevent the further spread of the COVID-19 and reduce the burden on healthcare providers and radiologists.
%
\bibliographystyle{IEEEbib}
\bibliography{refs}

\end{document}